# Comparative study of the binding energy in a thin and ultra-thin organic-inorganic perovskite within dielectric mismatches effects

Haitham ZAHRA,[1, a)] Aïda HICHRI,[1] Sihem JAZIRI[1,2]

[1]*Laboratoire de Physique des Matériaux, Faculté des Sciences de Bizerte, Université de Carthage, 7021 Jarzouna, Tunisia,*
[2]*Laboratoire de Physique de la Matière Condensée, Faculté des Sciences de Tunis, Université Tunis El Manar Campus Universitaire, 2092 Tunis, Tunisia*

a) haythemzahra20@gmail.com

**Abstract**

The multi-quantum well (MQW) organic-inorganic perovskite offer an approach of tuning the exciton binding energy based on the well-barrier dielectric mismatch effect, which called the image charge effect. The exfoliation from MQW organic-inorganic perovskite forms a two-dimensional (2D) nano-sheet. As with other 2D materials, like graphene or transition metal dichalcogenides (TMDs), the ultra-thin perovskites layers are highly sensitive to the dielectric environment. We investigate the ultrathin crystalline 2D van-der-Waals (vdW) layers of organic-inorganic perovskite crystals close to a surface of the substrate. We show that binding exciton energy is strongly influenced by the surrounding dielectric environment. We find that the Keldysh model somehow estimates the strong dependence of the exciton binding energies on environmental screening. We compare our binding energies results with experimental results in the $(C_6H_{13}NH_3)_2PbI_4$ perovskite, and we estimate the binding energy values of $(C_4H_9NH_3)_2PbBr_4$.

**I- Introduction**

For about a decade, 2D materials have represented one of the hottest directions in solid-state research. Much attention has been paid to 2D layered compounds such as graphene or TMDs. Due to weak vdW bonding, it is easy to cleave neighboring layers and form ultra-thin samples.[1-5] In these materials, new optical and electronic properties emerge for mono- or few-layer regions, providing new avenues for material applications. Here, we explore a recent addition to this library, ultrathin crystalline layers of organic-inorganic perovskite crystals. These materials differ from the previous types of 2D vdW layers in being a hybrid material with an organic compound intrinsically integrated into an inorganic crystal structure.

However, in order to be effectively integrated into vdW heterostructures with other 2D materials such as graphene and monolayer TMDs, the layers must be isolated in single-crystal form and be both atomically smooth and as thin as possible. In this context, the possibility of producing ultrathin organic-inorganic perovskites sheets via mechanical exfoliation was recently reported.[6]

Generally, perovskites represent a very large family of compounds, which ramify to many groups. One of them is the hybrid organic-inorganic perovskites which provide a significant opportunities as multifunctional materials for many electronic and optoelectronic applications, such as organic-inorganic field-effect transistors,[7] or nonlinear optical switches based on strong exciton-photon coupling in microcavity photonic architectures.[8] Very recently, hybrid organic-inorganic perovskites have been suggested as a new class of low-cost material for high efficiency photovoltaic cells. [9-12]

The layered organic-inorganic perovskite have been investigated as naturally occurring direct-gap MQW. The structure of layered perovskites consists of an alternation of inorganic and of organic layers.[12-17] The HOMO-LUMO (Highest Occupied Molecular Orbital and Lowest Unoccupied Molecular Orbital) energy gap of the organic layers is higher than the band gap of the inorganic layers (at least by 3 eV).[18] These two kinds of layers play as barriers and wells alternating with each other, inorganic layers are wells and the organic layers are barriers for the electron and the hole. Organic compounds offer a number of useful properties including structural diversity, ease of processing and high luminescence quantum yield at room temperature. On the other side, inorganic materials have a distinct set of advantages, including good electrical mobility, band gap tenability (enabling the design of metals, semiconductors, and insulators), mechanical and thermal stability, and interesting magnetic or electric properties.

For the mostly used layered $(R-NH_3)_2 MX_4$ perovskites, where R is an aliphatic or aromatic ammoniumcation, M is a divalent metal that can adopt an octahedral coordination, and X is a halogen: Cl, Br or I.[12] The width of the wells is mainly controlled by the M-X bond length, their depth is controlled by M and X species, and the width of the barrier is controlled by the organic radical R. It is thus possible to tune the sharp resonant emission wavelength in a large range from 320 to 800 nm by substituting different organic parts R, metal cations M or halides X.[19,20] For example, the perovskites containing Ge and Sn emit in the infrared range, perovskites containing Pb emit in the visible range.

In this paper, we focus our attention on a particular inorganic-organic QW crystals $(C_6H_{13}NH_3)_2PbI_4$ and $(C_4H_9NH_3)_2PbBr_4$, which are a self-organized crystals, in which the excitons are tightly confined in the inorganic layer of [PbI$_6$] ([PbBr$_6$]) octahedra sandwiched between organic barrier layers consisting of alkyl-ammonium chains [C$_6$H$_{13}$NH$_3$] ([C$_4$H$_9$NH$_3$]) respectively. This crystal has been attracting much interest because it exhibits many fascinating characteristics due to its unique crystal structure, such as huge optical nonlinearity with ultrafast response,[21] bright electroluminescence,[22] and outstanding scintillation characteristics.[23]

The excitonic effects in perovskite are not determined only by the quasi-particle confined energies, but also the Coulomb interaction between the electron and hole. The strength of the electron-hole interaction is characterized by the exciton binding energy E$_B$. The spatial electron and hole confinement in a very thin inorganic quantum well QW quadruples approximately the bulk exciton binding energy and halves its exciton Bohr radius a$_B$. However, perovskite materials present a dielectric constant in the barrier layer sizably smaller than that in the well layer ($\varepsilon_b < \varepsilon_w$). The theory suggests that the high contrast in the dielectric constants creates an enhancement of the binding energies.[24,25] This effect is called the dielectric confinement effect by analogy to the quantum confinement effect. It was predicts by Rytova[26] and Keldysh[27] who investigated excitons in a thin semiconductor film in dielectric surroundings. If we assume the 2D limit (well width is zero and barrier potential is infinity) of a single QW, the exciton binding energy relative to thethree-dimensional (3D) case will be enhanced as $E_{ex}^{2D} = 4\eta^2 E_{ex}^{3D}$,[27,28] where $E_{ex}^{3D}$ is the binding energy of the corresponding 3D excitons, the squared ratio $\eta^2 = (\frac{\varepsilon_w}{\varepsilon_b})^2$ of the dielectric constants results from the perfect image charge effect for a single QW.

Recently, similar excitonic effects have been studied theoretically and observed experimentally in atomically thin TMDs [29–41]. Although 2D materials TMDs are not directly bonded with the environment, due to their low thickness, they are highly sensitive to the dielectric screening of their surroundings. In particular, the Coulomb interaction between an electron and a hole in exciton is screened by the dielectric environment and the exciton binding energy changes dramatically.

Our analysis in this paper isinspired by an insightful paper of Yaffe et al [42] which deduce an excitonic binding energy of 490±30 meV for the ultrathin crystalline layers of organic-inorganic (C$_4$H$_9$NH$_3$)$_2$PbI$_4$. This binding energy exceeds the value for the corresponding layered quasi-2D material (370 meV) [43,44] and of course the cubic 3D systems (37 meV).[45]

Motivated by these results, and in order to understand the dielectric confinement and the dielectric environment effects of the 2D organic-inorganic nanosheet heterostructure, we propose to start in section II, to calculate by a theoretical model the binding exciton energy in the quasi-2D organic-inorganic perovskite using the potential image effect,[44, 46] which takes into account the contribution of the intrinsic variation of dielectric constantsbetween the well and the barrier. In section III, we study the organic-inorganic nanosheet, which is obtained by the exfoliation of the quasi-2D organic-inorganic perovskite onto SiO$_2$/Si substrate. Using Keldysh potential model, we show the sensibility of the thin layers with the dielectric environment (vacuum, SiO$_2$/Si) on the screening charges. [29, 42]

**II- Influence of the organic barrier in the organic-inorganic MQW:**

Let us consider an exciton in quasi-2D perovskites. The electronic structure can be regarded as a self-organized MQW in the growth direction along Oz. Assuming parabolic and isotropic band structure for the electron and the hole, the exciton wave function can be expressed by $\frac{1}{\sqrt{S}}e^{iK(\alpha_e r_e + \alpha_h r_h)}\psi(r_e, r_h)$, where S is the area of the layer, $K$ is the center of mass momentum, $r_{e(h)} = (\rho_{e(h)}, z_{e(h)})$ and $\alpha_{e(h)} = \frac{m^*_{e(h)}}{M}$ are the vector position and the mass ratio of the electron (hole) respectively, where $m^*_e$ and $m^*_h$ are the effective electron and hole masses respectively and M is the exciton effective mass which is given by $M = m^*_e + m^*_h$. In order to study the optical properties of the system, the center of mass momentum is conserved ($K$=0). The wave function $\psi(r_e, r_h)$ is the solution of the quasi 2D Hamiltonian which is expressed by $H_{ex} = -\frac{\hbar^2}{2m^*_e}\frac{\partial^2}{\partial z_e^2} + U^e(z_e) - \frac{\hbar^2}{2m^*_h}\frac{\partial^2}{\partial z_h^2} + U^h(z_h) - \frac{\hbar^2}{2\mu}\nabla_\rho^2 + V(z_e, z_h, |\rho_e - \rho_h|)$. The e-h potential energyis given in terms of the one electron (hole) potentials $U^{e(h)}(z_{e(h)})$ and the image potential e-h interaction $V(z_e, z_h, |\rho_e - \rho_h|)$.

The self-image potential $U^{e(h)}(z_{e(h)})$ will be composed of two term $U^{e(h)}(z_{e(h)}) = V^{c(v)} + \frac{e}{2}\int_0^\infty \frac{qdq}{2\pi}[\varphi(z_{e(h)}, z_{e(h)}, q) - \frac{2\pi e}{\varepsilon(z_{e(h)})q}]$ where q=|$\mathbf{q}$|, $\mathbf{q}$ is the in-plan wave vector (see ref 20), $V^{c(v)}$ is the conduction (valence) band offset, $\varphi(z_{e(h)}, z_{e(h)}, q)$ is the electrostatic potential.The effect of the dielectric-constant difference between the well and barrier layers is included in the electrostatic potential. The image-charge method, which is a well-established method in electrostatics, represents the electric field induced by charged particles in the plane parallel geometries in terms of imaginary charges placed in virtually homogeneous media, which is derived by R. Guseinov. [46] To study the exciton state, we firstly have to evaluate the single electron (hole) confinement energy along Oz direction. It can be calculated through the Schrodinger equation for motion perpendicular to theperovskite layers, which corresponds to the one-electron( one-hole) self-image potential $U^{e,h}(z_{e(h)})$. The one electron and one-hole Schrodinger equations can be written as

$$\left(\frac{-\hbar^2}{2\,m^*_{e(h)}}\frac{\partial^2}{\partial z^2_{e(h)}} + U^{e(h)}(z_{e(h)})\right)\Phi^{e(h)}(z_{e(h)}) = E^{e(h)}\Phi^{e(h)}(z_{e(h)})$$

The term of the relative quasi- 2D exciton Hamiltonian is given as follows $H_{rel}^{Quasi-2D} = -\frac{\hbar^2}{2\mu}\nabla_\rho^2 + V(z_e, z_h, |\boldsymbol{\rho}_e - \boldsymbol{\rho}_h|)$, where $\mu = \frac{m^*_e m^*_h}{m^*_e + m^*_h}$ is the reduced effective mass. The Coulomb interaction potential $V(z_e, z_h, |\boldsymbol{\rho}_e - \boldsymbol{\rho}_h|)$ will be averaged by the probabilities of the electron and the hole presence probabilities along the Oz axis. In order to study the effect of the dielectric-constant difference in a more realistic situation, we examine two materials $(C_4H_9NH_3)_2PbBr_4$ and $(C_6H_{13}NH_3)_2PbI_4$ denoted as $C_4PBr$ and $C_6PI$ respectivelywith characteristic parameters defined in table I below:

The $C_4PBr$ and $C_6PI$ have the same dielectric constants of the barrier layers ($\varepsilon_b = 2.1$) but the dielectric constants of the well layers are different ($C_4PBr$: $\varepsilon_w = 4.8$; $C_6PI$: $\varepsilon_w = 6.1$). The remarkably difference of the image effect between $C_4PBr$ and $C_6PI$ mainly originates the smaller dielectric ratio η of the $C_4PBr$ and the high η of $C_6PI$ ($\eta_{C_6PBr} = 2.28$ and $\eta_{C_6PI} = 2.9$). On the other hand, to say that this effect is mainly due to the dielectric constant, the calculated Bohr radius of this materials is slightly different (between 12 and 14Å), this is reasonable because of the nearly equal reduced effective mass ($\mu_{C_6PI}= 0.18m_0$ and $\mu_{C_4PBr}= 0.17m_0$).

In this work we are mainly interested in the effect of dielectric constants. However, the effective mass of the carriers is important too, because introducing the effective mass mismatch between the organic and inorganic layers is equivalent to adding contributions to the confinement potentials. Following Even et al.,[48] we take the same effective masses for both particles (electron, hole) in the same region (QW or the barrier) but different between the QW and the barrier. Moreover, real thickness $l_{w(b)}$ of the QW (the barrier) can be considered as an adjustable parameter and may differ from values taken in table I due to the nature of the QW–barrier interface.[24] The reference 3D material $(CH_3NH_3)\,PbI_3$ [49] was used for the value of $E_g^{3D}$.

In Fig. 1, we plot the image potentialof $C_4PBr$ and $C_6PI$ along Oz direction. The interaction between the induced image charge and the original chargeis solved numerically. Fig. 1 shows a divergence close to the interface due to the high contrast of the dielectric constants. This effect causes an increase of the particle energies and thus a blue-shift of the electron-hole band gap transitions. The physical reason for this is the repulsion of QW localized electrons and holes from the interfaces. However, we cannot leave the divergence, which are unphysical and appear in our model as a result of the unrealistic assumption of local dielectric susceptibility changing abruptly at the interfaces. To solve this problem we add a transitional layer width which is an additional adjustable parameter Δ with the used material (in the order of interatomic distance) (see fig.2).

The numerical resolution of the one-electron and one-hole Schrodinger equations enables us to calculate the confinement energy in perpendicular motion for electron and hole. However, one additional effect so far has not been taken into account and stems from the fact that in the perovskite films the quantum wells are not separate but are stacked together. The electron and hole wave functions, which extend outside of the quantum wells, can thus hybridize with those from neighboring quantum wells QWs. In the other hand, the energy band gap of the organic layers is higher than the band gap of the inorganic layers (more than 3eV), which makes the carriers strongly confined in the well and the effect of the nearest well is weak.The large band offsets allow us to predict a low tunneling between neighboring QW layers (see figure 2). We find that the Tight-binding model can estimate this effect. The wave function takes the form of Bloch function $\frac{1}{\sqrt{N}}\sum_p e^{ikpd}\Phi^{e(h)}(z_{e(h)} - pd)$ where d= $(l_w + l_b)$ is the period of the MQW and **k** is the wave vector along Oz direction.

In this paper, and to avoid tedious calculation of the quantization energies of the electron(hole) $E^{e(h)}$ respectively, we have in one hand approximated that the electron and hole are confined in asquare potential.We used $l_w$ as adjustable parameters under the condition $l_w + l_b = (5.9 + 8.1)$ Å for($C_4H_9NH_3)_2PbBr_4$ and $l_w + l_b = (6.3 + 10.03)$Å for ($C_6H_{13}NH_3)_2PbI_4$) respectively. In spite of this rather crude simplification, the obtained results satisfactorily reproduce the experimental results [49] for example in ($C_6H_{13}NH_3)_2PbI_4$ the ground eigenvalues are $E^e$=0.62 eV and $E^h$=0.41eV.

Now, to construct the complete exciton potential, we have to solve the excitonic Schrodinger equation, for the in-plane motion, corresponding to the averaged image-potential-mediated e-h interactionV (ρ).We calculated the *ns* exciton binding energies $E_n^b$ based on Muljarov's et al. formalism.[24] The relative Hamiltonian of the system within the effective mass approximation is given by $H_{rel}^{Quasi-2D} = -\frac{\hbar^2 \boldsymbol{\nabla}_\rho^2}{2\mu} + V(\rho)$, where $V(\rho)$ is the averaged image-charge-mediated potential $V(\rho) = \int dz_e \int dz_h \, |\Phi^e(z_e)|^2 |\Phi^h(z_h)|^2 V(z_e, z_h, \rho)$.The relative Hamiltonian can be written as a summation of two Hamiltonians, where the solution of the

first one $H_0 = -\frac{\hbar^2 \nabla_\rho^2}{2\mu} - \frac{e^2}{\varepsilon^* \rho}$ is known and exact, corresponds to the 2D hydrogenic states. The exciton spectrum of $H_0$ is given by $E_n = -\frac{R^*}{\left(n-\frac{1}{2}\right)^2}$ where $R^* = \frac{e^4 \mu}{2\varepsilon^{*2}\hbar^2} = \frac{e^2}{2\varepsilon^* a^*}$ is the three dimensions effective Rydberg and $\varepsilon^*$ is the average dielectric constant, it can be written as $\varepsilon^* = \sqrt{\varepsilon_w \varepsilon_b \frac{(\varepsilon_w l_w + \varepsilon_b l_b)}{(\varepsilon_b l_w + \varepsilon_w l_b)}}$ [24] and $a^* = \frac{\varepsilon^* \hbar^2}{\mu e^2}$ is the exciton effective Bohr radius. It should be noted that the effective dielectric constants $\varepsilon^*$ is appropriate if the material is assumed to be uniform continuous matter, therefore an estimation for the enhancement of the image charge effect using $\varepsilon^*$ leads to an underestimation. The second term $H_1 = \frac{e^2}{\varepsilon^* \rho} + V(\rho)$ can be considered as a perturbation. The enhancement of the Coulomb interaction in a thin inorganic semiconductor layers sandwiched by organic layers is caused by the effective reduction of the dielectric constants. Using numerical diagonalization of the Hamiltonian $H_{rel}^{Quasi-2D}$, we obtain the eigenvalues $E_{rel}^{2D}$ and the eigenstates $\phi(\rho)$ which are expanded in terms of the hydrogenic wave functions as $\phi(\rho) = \sum_{n,m} C_{n,m} a_{n,m} e^{il\theta} \rho^{|m|} e^{-\frac{\rho}{2}} L_{n-|m|-1}^{2|m|}(\rho)$, [50,51] $n = 1, 2, 3 \ldots$ with angular momentum $m = 0, \pm 1, \pm 2, \pm 3, \ldots \pm n - 1$. The states are ($2n$-1) fold degenerate, the states are labeled S for $m=0$, $p$ for $m=\pm 1$ and $d$ for $m=\pm 2$. In order to compare our results to experimental findings, we have limited ourselves to the 5S states. The excitonic binding energy is given by $E_{nS}^b = -E_{rel}^{2D}$.

Fig. 3a shows the in-plane image potential of C$_4$PBr and C$_6$PI. The image potential of the C$_4$PBr appears to be more localizing in comparison with that of the C$_6$PI due mainly to the dielectric mismatch effect between the well and the barrier. In Fig. 3b, we plotted the calculated $\rho$ dependence of the image potential $V(\rho)$ of the C$_6$PI solid lined and the dotted lines represent the ideal 2D Coulomb potential with a perfect image charge enhancement $V^{2D}(\rho)$ (the potential $V^{2D}(\rho)$ is given by $V^{2D}(\rho) = -e^2/\varepsilon^* \rho$).[24] $V(\rho)$ agrees well with the ideal 2D Coulomb $V^{2D}(\rho)$ potential in the range $\rho > 12$ Å, it follows that the nS (n≥ 2) excitons form an ideal 2D hydrogenic series. The radial probability density of C$_6$PI 1S exciton is presented in Fig.3b by a dashed lines and it occupies the inner range of $\rho \leq 12$ Å, with the 2D hydrogenic model, the radial probability density would be maximum in $\frac{a^*}{4} = 3.42$ Å which is larger than inferred for that calculated with image effect giving 3.28Å (see fig 3b). It should be noted also and for comparison, that the 1S exciton binding energy is calculated to be 80 meV when the image charge effect is not included ($\varepsilon_w = \varepsilon_b = 6.1$), which is at least 2.16 times larger as the binding energy of the exciton in the 3D analogue (bulk crystal), ($CH_3NH_3PI_3$) (37 meV).[45] Since, the spatial confinement enhances the 1S exciton quadruplet the exciton binding energy in the 2D limit. The 2.16 enhancement factor indicates that the spatial confinement for the 1Sexciton is not sufficient.Nevertheless, fig. 3b shows the obtained exciton binding energies $E_{nS}^b$ of $C_6PI$, which are equal to $E_{1S}^b$= 373 meV, $E_{2S}^b$= 55 meV and $E_{3S}^b$= 32 meV and presented by arrows. It is worth noting that the 1S excitons binding energy (373 meV) is still at least 10 times larger than that in the 3D analogue. The additional enhancement by a factor of 10/2.16=4.6 is definitelydue to the image charge effect. The obtained results satisfactorily reproduce the Tanaka results, as will be shown in the inset of Fig. 3b.

**III-    influence of the dielectric environment onthe screening of charges within the ultrathin organic-inorganic perovskite crystals**

From the previous paragraph, we deduce that the binding energy of $(C_4H_9NH_3)_2PbBr_4$ (($C_6H_{13}NH_3)_2PbI_4$) is equal to 390 meV (373meV). This binding energy exceeds the value for the corresponding of the cubic 3D organic-inorganic systems (37meV).[45] Omer Yaffe et al. show that the binding energy is 490 ±30 meV in ultrathin perovskite PbI-based.[42] This energy is obtained from large sheets of layered organic-inorganic perovskite crystals, as thin as a single unit cell. Typical layers produced by the exfoliation of bulk of the perovskite crystals onto $SiO_2$/Si substrates are found to have lateral dimensions of tens of microns.[42] The increase of the binding energy is strongly influenced by the effect of quantum confinement and the dielectric environment. Quantum confinement restricts the spatial extent of the exciton wave function in the perpendicular direction. An additional contribution to the excitonic binding energy arises from the no uniform dielectric environment, since the electric field between the electron and hole forming an exciton extendsoutside the inorganic layer into the surrounding medium.[27,37,40]

The nonlocally-screened electron-hole interaction due to the screening caused by the change in the dielectric environment is typically described by an effective dielectric constant. To estimate the binding energies of the exciton states in organic inorganic perovskite, as thin as a single unit cellwe consider the relative exciton Hamiltonian $H_{rel}^{2D} = -\frac{\hbar^2 \nabla_\rho^2}{2\mu} - \frac{2\pi e^2}{(\varepsilon_1 + \varepsilon_2) r_s} [H_0(\frac{\rho}{r_s}) - Y_0(\frac{\rho}{r_s})]$.[29,31,40] The second term of $H_{rel}^{2D}$ is a nonlocally-screened electron-hole interaction due to the screening caused by the change in the dielectric environment, where $H_0(x)$ and $Y_0(x)$ are Struve and Neumann functions, respectively,$\varepsilon_1$ (air) and $\varepsilon_2$ (substrate) are the environmental relative dielectric constants, surrounding the perovskite layer. The organic/inorganic perovskite thin layer is on the top of $SiO_2$/Si substrate characterized by $\varepsilon_2 =3.9$,[43] while the top surface is exposed to the air $\varepsilon_1 =1$, and $r_s = \frac{\varepsilon^* D}{\varepsilon_1+\varepsilon_2}$ [28,30] is the screening factor, where $D$ and $\varepsilon^*$ are the effective width of the ultra-thin layer [6,38] and the effective dielectric constant respectively. In fact, the screening length incorporates the experimental environment and is highly sensitive to electromagnetic fields, doping, or dielectric screening of surrounding materials. Lin et al. [52] and Li et al. [53] deepened the study of the influence of the dielectric constant of the surrounding environments on the exciton behaviors in $MoS_2$ mono layer. We restrict ourselves to the low lying nS states (n=1,2,..5) which depend strongly on the screening length $r_s$. The screening length $r_s$ is equal to 12 Å for $(C_6H_{13}NH_3)_2PbI_4$ (10.5 Å for $(C_4H_9NH_3)_2PbBr_4$). The fig. 4a shows that Keldysh potentials of $(C_4H_9NH_3)_2PbBr_4$ (red line) and $(C_6H_{13}NH_3)_2PbI_4$ (black line) are more localized in comparison with that calculated without environment dielectric effect (dotted lines) and are slightly different for each perovskite ultra-thin layer. This reflects the influence of the dielectric environment on the screening of charges in the perovskite ultra-thin layer. In Fig. 4b, we plot the calculated ρ dependence of the nonlocally-screened potential V(ρ) of the $(C_6H_{13}NH_3)_2PbI_4$ indicated by a solid lineand the one without the environment corrections $V^{2D}(\rho)$ represented with the dotted lines. The 1S exciton radial probability density is presented in fig. 4b by a dashed lines and it occupies the inner range of $\rho \leq 15$Å. Its maximum shifts in the same sense as with the image effect but more importantly and situated at 2.8Å. This figure shows that the change in the dielectric environment of these ultrathin perovskite layersleads to an increase in the exciton binding energy about 1/3 greater than the corresponding quasi-2D materials (as will be shown in the table II). This is mainly due to the field produced by these charges in the dielectric environment surrounding the monolayer, which begins to play a perceptible role. The inset of Fig. 4b shows that our results are slightly different with the experimental results of ref.[42] This implies that the screened Coulomb potential in a 2D system given by L.V. Keldysh, can explain the behavior of exciton in the 2D ultra-thin perovskite.

For more general analysis of our system, we can predict the binding energies of the perovskite thin layer by changing the surrounding environment. Fig. 5 shows the calculated exciton binding energy of $(C_6H_{13}NH_3)_2PbI_4$ thin layer, encapsulated between two thick layers of varying dielectric constants. As we show, fora low values of the environment dielectric constants, the binding energy is higher than 675meV. This figure shows also, that when the perovskite and the environment have the same order of magnitude of the dielectric constant, the surrounding has no effect on the perovskite layer. More generally, the interaction between charge carriers is highly sensitive to the local dielectric environment. Correspondingly, the exciton binding energy is expected to be highly tunable by means of a deliberate change of this environment, as illustrated in Fig. 5, like the influence of a solvent on the perovskite ultra-thin layer. We can conclude that the ultra-thin 2D perovskite can offer a new approach for tuning the energies of the electronic states based on the interplay between the environmental sensitively and unusual strength of the Coulomb interaction in these materials.

## IV- Conclusion

We have studied the influence of the dielectric constant mismatch between organic-inorganic layers in the MQW perovskite in the first case, and between the ultra-thin perovskite layer and the surrounding environment in the second case for the $(C_4H_9NH_3)_2PbBr_4$ and $(C_6H_{13}NH_3)_2PbI_4$ materials. We show that the organic layer enhances the binding energy in the quasi-2D perovskite layer due to the effective dielectric constant. We show also that in the ultra-thin layer (2D perovskite) the binding energy is strongly influenced by the surrounding dielectric environment. We remark that the 2D exciton binding energies for the $(C_4H_9NH_3)_2PbBr_4$ and $(C_6H_{13}NH_3)_2PbI_4$ (equal to 482 and 470 meV respectively) are slightly different. In the other words, this is can be explained by the important effect of the surroundings on the exciton 2D binding energy in comparison with the 3D exciton or bulk binding energy which are very different for the two kind of materials.

**Figure and table captions:**

**Figure 1:** Image potential along Oz direction in $(C_4H_9NH_3)_2PbBr_4$ (red line) and $(C_6H_{13}NH_3)_2PbI_4$ (black line). This figure presents the image potential between the induced image charge and the original charge (electron or hole) for the different values of the ratio η in the conduction band.

**Figure 2:** Representation of the electron (black line) and the hole (red line) potentials in the $(C_6H_{13}NH_3)_2PbI_4$ used to compute the fundamental energies of the carriers. A separated quantum well with the presence of image effect, the interface Δ and the corresponding wave functions.

**Figure 3:** a) Image-potential-mediated electron-hole interaction V(ρ) (thick solid line) of $(C_4H_9NH_3)_2PbBr_4$ (red line) and $(C_6H_{13}NH_3)_2PbI_4$ (black line). b) A plot of the 2D radial probability density ρ|ϕ(ρ)|² (dashed line), where $\phi(\rho)$ is the 1S exciton wave function which is computed by numerically solving the exciton Hamiltonian, including the image potential V(ρ) ( Thick solid line), electron-hole interaction without image corrections $V^{2D}(\rho)$ (dotted line) and the corresponding energies are shown by thin solid arrow. **Inset figure:** Exciton binding energies for the states (1S, 2S and 3S) obtained from Tanaka et al. results [43] (red dots) are compared with our theoretical model (black dots).

**Figure 4:** a) Keldysh electron-hole interaction V(ρ) of $(C_4H_9NH_3)_2PbBr_4$ (red line) and $(C_6H_{13}NH_3)_2PbI_4$ (black line) with (thick solid line), and without environment corrections (dotted line) $V^{2D}(\rho)$. b) A plot of the 2D radial probability density ρ|ϕ(ρ)|² (dashed line) where $\phi_{1S}(\rho)$ is the 1S exciton wave function which is computed by numerically solving the exciton Hamiltonian, including the non-local potential V(ρ) ( Thick solid line), electron-hole interaction without environment corrections $V^{2D}(\rho)$ (dotted line) and the corresponding energies are shown by thin solid arrow. **Inset figure:** Exciton binding energies for the states (1S, 2S and 3S) obtained from experimental results [42] (red dots) are compared with our theoretical model (black dots).

**Figure 5:** Influence of the surrounding dielectric environment on the exciton binging energy of $(C_6H_{13}NH_3)_2PbI_4$. The figure shows an overview of predicted changes in the exciton binding energy in $(C_6H_{13}NH_3)_2PbI_4$, encapsulated between two thick layers of dielectrics constants ($\varepsilon_{Top}$ and $\varepsilon_{Bottom}$).

**TABLE I.** Parameters of the perovskites materials.

**TABLE II.** Comparison of the exciton binding energies in $(C_4H_9NH_3)_2PbBr_4$ and $(C_6H_{13}NH_3)_2PbI_4$.

**FIGURES:**

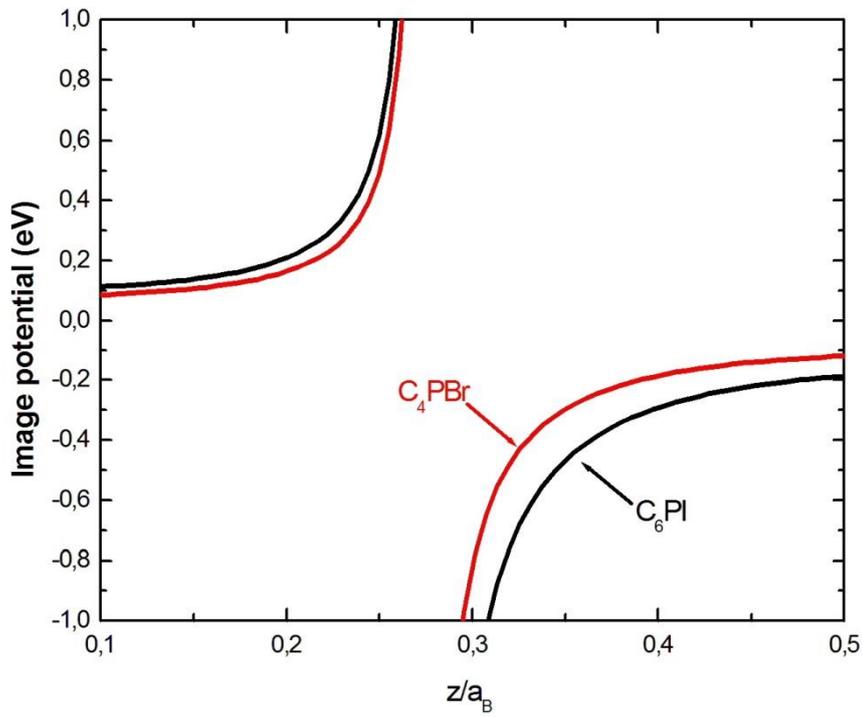

FIG.1

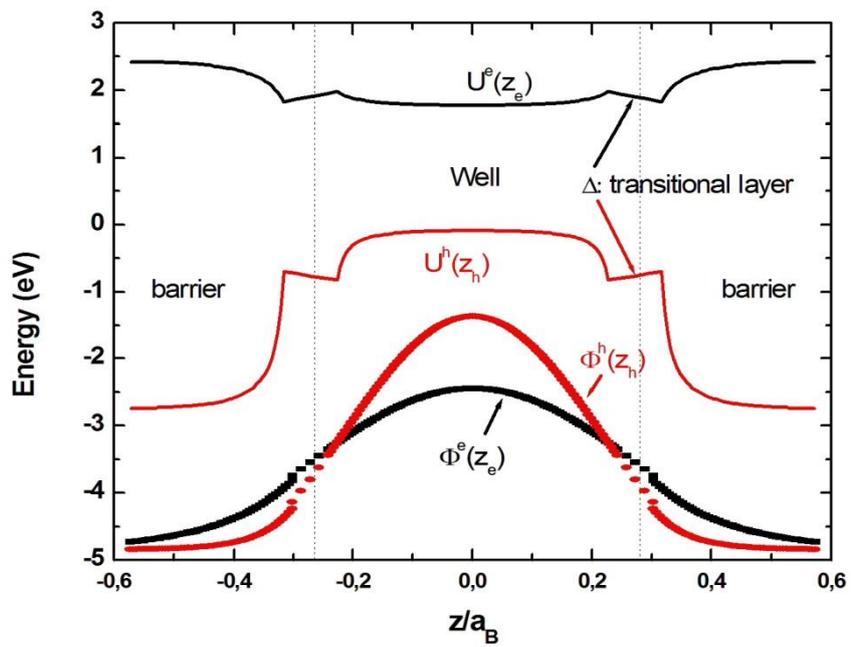

FIG.2

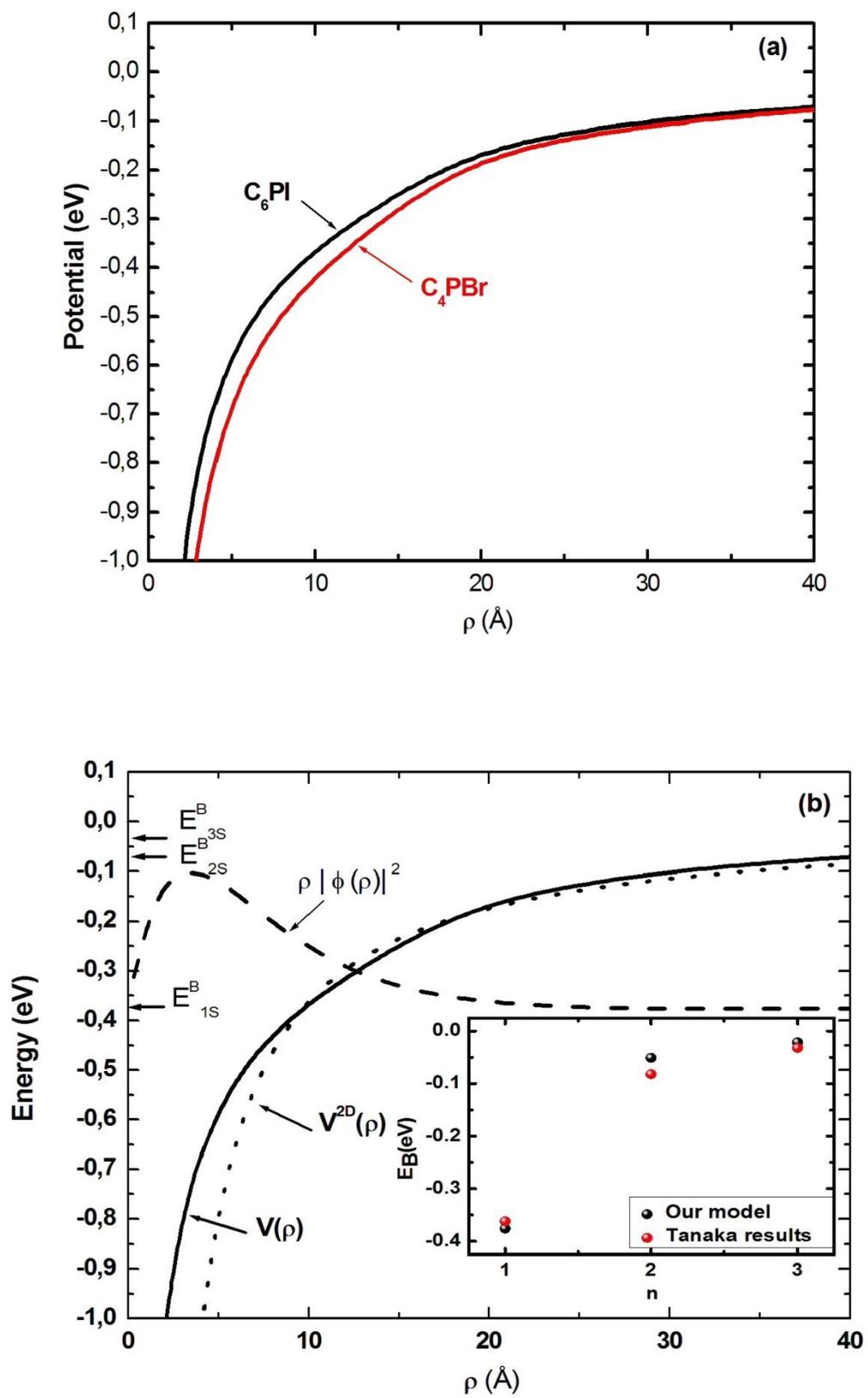

FIG.3

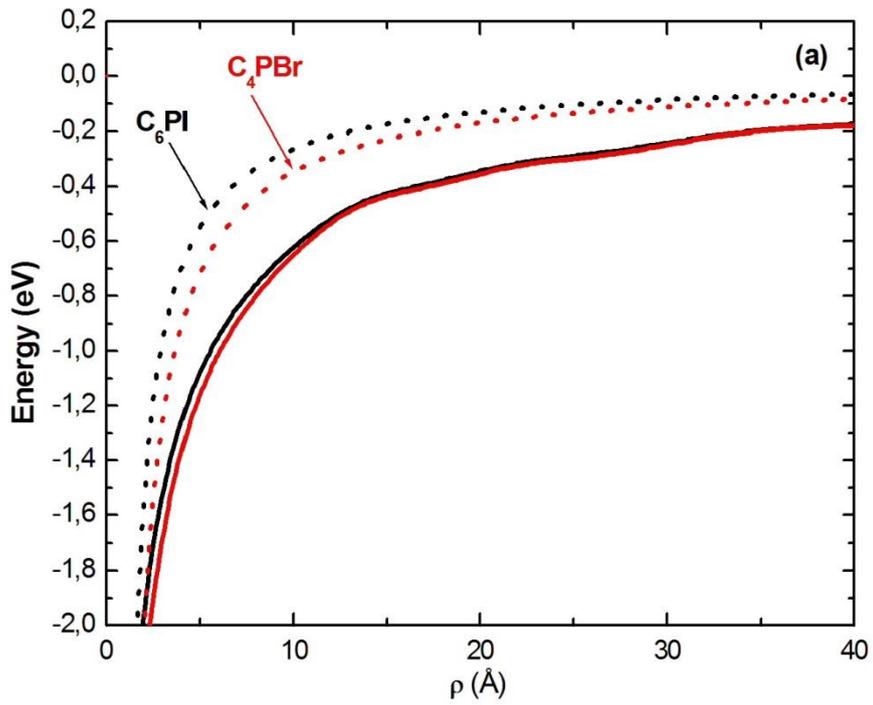

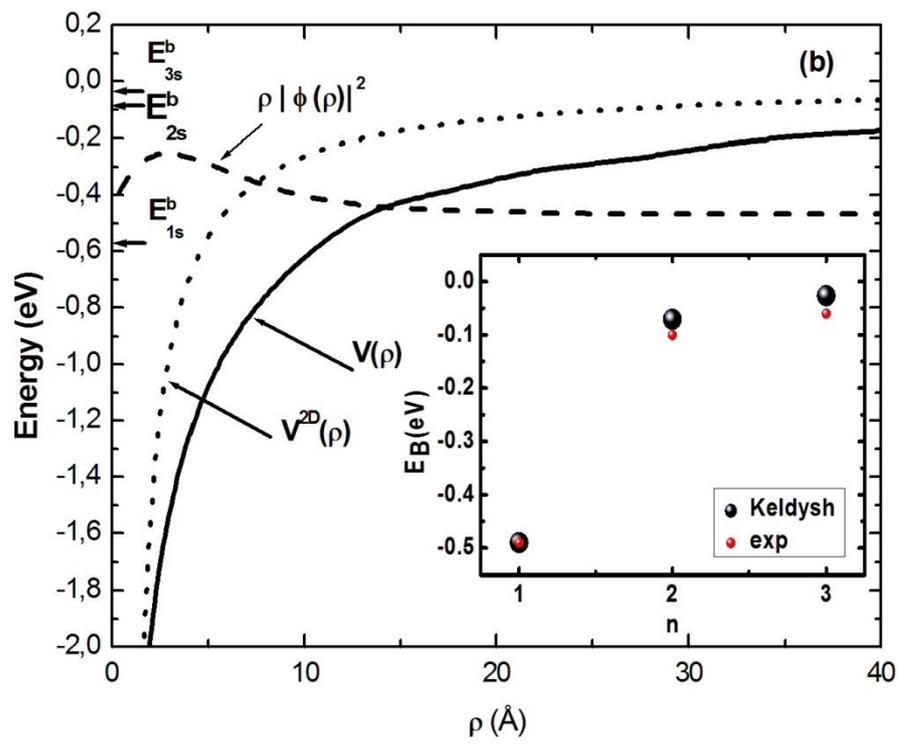

FIG.4

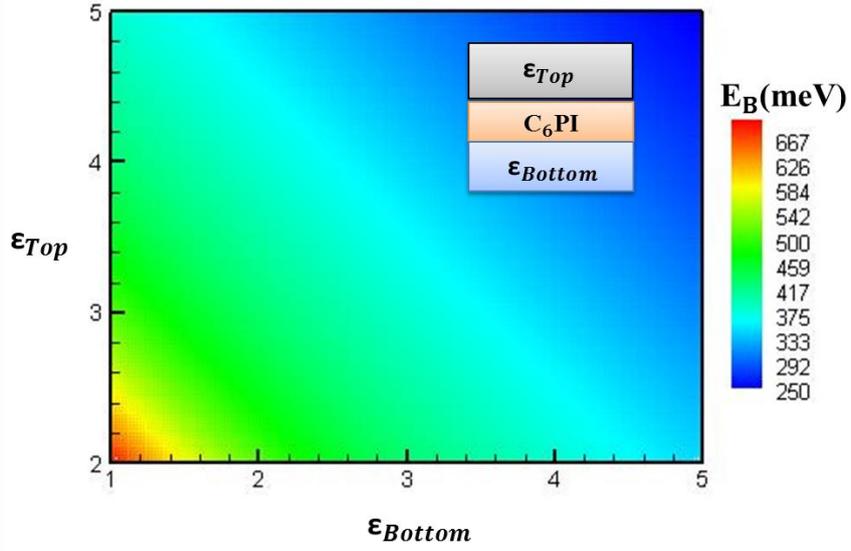

FIG.5

|  | $\varepsilon_w$ | $\varepsilon_b$ | $l_w$(Å) | $l_b$(Å) | η | μ | $a_B$(Å) |
|---|---|---|---|---|---|---|---|
| $(C_4H_9NH_3)_2PbBr_4$ | 4.8[a] | 2.1-2.4[a,b,c] | 5.9[c] | 8.1[c] | 2.28 | 0.17[c] | 12[c] |
| $(C_6H_{13}NH_3)_2PbI_4$ | 6.1[b] | 2.1[b] | 6.3[b] | 10.03[b] | 2.9 | 0.18[b,c] | 14[b,c] |

[a]Reference 28

[b]Reference 24

[c]Reference 47

TABLE I.

|  | $E_B^{3D}$ (meV) | $E_B$ in the MQW (meV) | | $E_B$ in the ultra-thin perovskite layer (meV) | |
|---|---|---|---|---|---|
|  |  | Theoretical results | Experimental results | Theoretical results | Experimental results |
| $(C_4H_9NH_3)_2PbBr_4$ | 76[d] | 390 | 393[e] | 482 |  |
| $(C_6H_{13}NH_3)_2PbI_4$ | 37[d] | 373 | 361[f] | 470 | 490± 30[g] |

[d]Reference 18

[e]Reference 28

[f]Reference 43

[g]Reference 42

TABLE II.